\documentclass[english,twocolumn,aps,prb,superscriptaddress,]{revtex4-2}
\usepackage{babel}
\usepackage{amsmath,mathtools}
\usepackage{amssymb}
\usepackage{graphicx}
\usepackage{tabularx}
\usepackage{lipsum}
\usepackage{xcolor}
\begin{document}
\title{NHSE-Driven Coalescence of Topological Defect States in Non-Hermitian Systems}
\author{S. M. Rafi-Ul-Islam}
\email{rafiul.islam@u.nus.edu}
\affiliation{Department of Electrical and Computer Engineering, National University of Singapore, Singapore 117583, Republic of Singapore}
\author{Zhuo Bin Siu}
\email{elesiuz@nus.edu.sg}
\affiliation{Department of Electrical and Computer Engineering, National University of Singapore, Singapore 117583, Republic of Singapore}
\author{Md. Saddam Hossain Razo}
\email{shrazo@u.nus.edu}
\affiliation{Department of Electrical and Computer Engineering, National University of Singapore, Singapore 117583, Republic of Singapore}
%\author{Haydar Sahin}
%\email{sahinhaydar@u.nus.edu}
%\affiliation{Department of Electrical and Computer Engineering, National University of Singapore, Singapore 117583, Republic of Singapore}
%\affiliation{Institute of High Performance Computing, A*STAR, Singapore 138632, Republic of Singapore}
\author{Mansoor B.A. Jalil}
\email{elembaj@nus.edu.sg}
\affiliation{Department of Electrical and Computer Engineering, National University of Singapore, Singapore 117583, Republic of Singapore}
\begin{abstract}
In this work, we describe a novel localization phenomena, the so-called ``topological defect accumulation", occurring in a non-Hermitian chain with an arbitrary number of defect sites. Specifically, it refers to the localization and coalescence of multiple defect eigenstates at a single defect site closest to the localization edge of the bulk   non-Hermitian skin modes. This phenomenon is distinct from the conventional topological defect states in Hermitian systems, where the defect states are separately localized at their respective defect sites.  The requirement for the onset of topological defect accumulation is the presence of non-reciprocal coupling which renders the chain non-Hermitian, as well as imaginary onsite potentials at the defect nodes. This allows the defect state accumulation and distribution to be 
modulated by both the defect site distribution and their corresponding onsite potentials paving the way for possible applications. Furthermore, this phenomenon is realizable in various synthetic and experimentally realizable platforms, such as topolectrical and photonic systems.
\end{abstract}

\maketitle
\section{Introduction}
In recent years, the study of non-Hermitian systems \cite{ashida2020non,rafi2021non,Siu_2025,rafiulislam2025topologicalzeromodesnonhermitian,bergholtz2021exceptional,gong2018topological,rafi2024knots,leykam2017edge,rafi2020topoelectrical,rafi2019transport} has received significant attention owing to their fascinating and unique properties \cite{gu2022transient,rafiulislam2024exceptionalpointsbraidingtopology,rafi2024saturation,gu2021controlling,2024APS..MARN17003R,siu2022critical,longhi2022self,2024APS..MARJ00258R,gu2021controlling} that are not observed in Hermitian systems \cite{rafi2020strain,sun2020spin,sun2019field,xu2023photoelectric,rafi2023conductance,rafi2022valley}. The field of non-Hermitian physics is fast-expanding and encompasses various different fields \cite{rafi2015efficient,islam2014thermal} and material systems, including condensed matter physics \cite{martinez2018topological,sahin2024protected,sahin2022interfacial,lee2019topological,rafi2023valley,sahin2022unconventional,zhang2022universal}, optics \cite{el2019dawn,wang2021topological}, photonics \cite{feng2017non,wang2021topological,rafi2022type,pan2018photonic,longhi2018parity}, topolectrical circuits \cite{helbig2020generalized,zhang2023anomalous,rafi2020anti,rafi2024chiral,rafi2020realization,rafi2022system,rafi2021non,rafi2022unconventional,zhang2023electrical,rafi2021topological,liu2020gain,rafi2022interfacial,rafi2022system}, and quantum mechanics \cite{hatano1996localization,hatano1997vortex,longhi2010optical}. Non-Hermitian systems generally exhibit the non-Hermitian skin effect (NHSE)\cite{okuma2020topological,rafi2025critical,rafi2024dynamic,rafi2022critical,li2020critical,longhi2019probing,song2019non}\textcolor{red}{,\cite{zhang2022review,lin2023topological}}, which refers to the exponential localization of eigenstates in the vicinity of an open boundary of the system. Additionally, recent studies have explored the interplay between topology and non-Hermiticity \cite{martinez2018topological,yao2018edge,yuce2018edge}\textcolor{red}{,\cite{yokomizo2021non, roy2025topological, zhang2020non, yokomizo2019non, zhou2018non, halder2024parsing,cao2021non}}. The non-trivial topology of a system can be characterized mathematically by means of topological invariants such as the Chern \cite{yao2018non,fujita2011gauge,shen2018topological} or winding number \cite{zhang2020correspondence,zeng2020winding,claes2021skin}. In Hermitian systems, the presence of topological states is related to the topology of the band structure, which can be described using the conventional concept of the Brillouin zone. However, in non-Hermitian systems, the characterization of its underlying topology requires the generalized Brillouin zone (GBZ) theory \cite{yang2020non,song2019non}.

Topological defect states are significant in topological physics because they are robust against perturbations and arise from changes in topological invariants across defects, enabling applications in quantum devices \cite{takei2018quantum,rafiulislam2025reconfigurabledefectstatesnonhermitian,chai2022topological}. In non-Hermitian systems, these states differ fundamentally from Hermitian cases: while Hermitian defect states localize separately at each defect, non-Hermitian ones coalesce at the defect nearest the NHSE edge due to non-reciprocal coupling, offering tunable clustering \cite{rafi2025frequency,rafiulislam2024anomalousnonhermitianskineffects,rafiulislam2019transporttopoelectricalweylsemimetal,rafiulislam2024engineeringhigherorderdiracweyl}. Studying this coalescence is important conceptually, as it reveals NHSE-topology interplay, and practically, for enhanced localization control in sensing and mode selection. This connects to broader non-Hermitian goals, such as engineering skin modes for signal amplification or robust information storage.

Another fascinating phenomenon in non-Hermitian systems is the emergence of robust topological defect states \cite{blanco2016topological,teo2017topological,stegmaier2021topological} upon the introduction of defects to an otherwise perfectly periodic system. In general, topological defects are regions where the order parameter (such as the magnetization or the density) changes abruptly. Topological defect states are states that are induced by these defects and are localized in their vicinity. The origin of topological defect states is linked to the topology of the system, and they are characterized by their energy and spatial distribution. Topological defect states have been observed in many diverse physical systems, such as liquid crystals \cite{lavrentovich1998topological,toth2002hydrodynamics}, superconductors \cite{teo2010topological,frolov2020topological}, and magnetic materials \cite{sumiyoshi2016torsional}. Their robustness against perturbations make them a promising candidate for novel quantum devices and technologies \cite{takei2018quantum,chai2022topological}.

\begin{figure*}[ht!]
  \centering
    \includegraphics[width=0.8\textwidth]{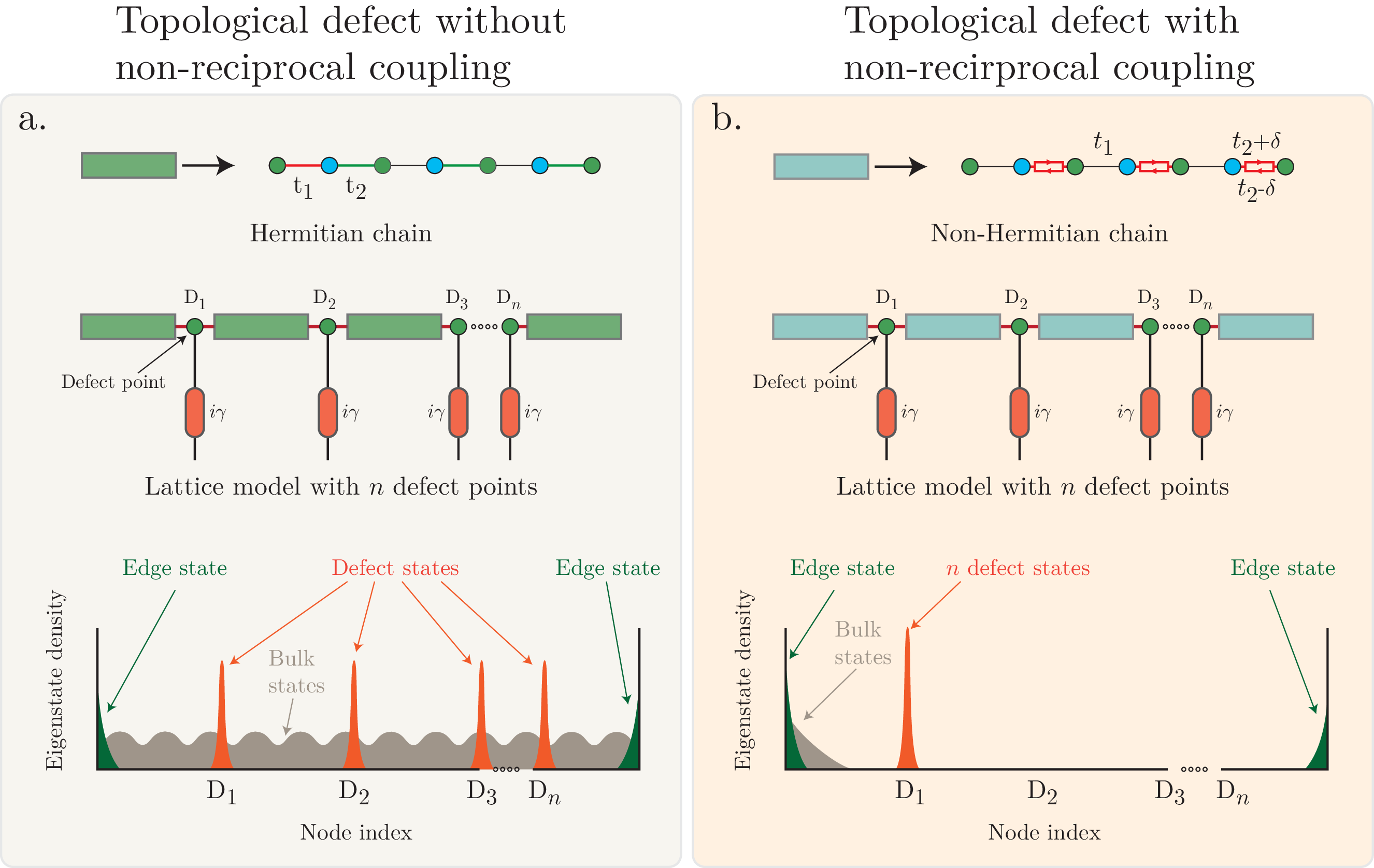}
  \caption{Non-Hermitian topological defect accumulation phenomenon in a chain with arbitrary number of defect sites. (a) Schematic illustration of conventional topological defect states that occur in the Hermitian limit, where all the defect states appear at their respective defect points. (b) Schematic of the novel topological defect accumulation phenomenon, where the eigenstates corresponding to the defect states accumulate at only one of the defect sites near the localization edge of the NHSE bulk modes. This phenomenon is distinct from the NHSE and edge state localization. The clustering of the defect states at only one of the defect sites depends on the complex energy and coupling asymmetry in the original chain.}
  \label{fig1}
\end{figure*}

In this study, we study the interplay between topological defect states and the NHSE, which gives rise to the so-called ``topological defect accumulation" in non-Hermitian chains with an arbitrary number of defect sites. Specifically, we observe the accumulation and coalescence / clustering of multiple \textcolor{red}{states} at a single defect site that is closest to the NHSE localization edge of the bulk states.  This accumulation is in contrast to that of the Hermitian system where each  defect state is  separately localized at its own respective defect site in the lattice chain (Fig. \ref{fig1}a).  The non-Hermitian topological defect accumulation  is characterized by a coalescing or clustering of the defect states, which is distinct from the conventional NHSE bulk states localization. As shown in Fig. \ref{fig1}b, this phenomenon  generally occurs whenever a non-Hermitian system with non-reciprocal coupling hosts arbitrary defect sites with dissimilar onsite potentials to that of the bulk nodes. By modulating the defect site distribution, its on-site potential or the degree of non-reciprocity of coupling in the chain, one can achieve a high degree of flexibility and tunability to the non-Hermitian topological defect accumulation. Furthermore, the non-Hermitian chains which host the topological defect states can be readily implemented in various synthetic platforms such as topolectrical and photonic systems. Given their tunability and robustness, as well as their ready implementation, these topological defect states can potentially be utilized in applications such as sensing and information storage.

The manuscript is organized as follows: Section II presents results on defect accumulation in uniform, alternating, and dissimilar gain/loss cases, including a new subsection on disorder effects. Section III discusses significance and applications. Section IV concludes, with Supplementary Materials providing derivations and additional figures.

\section{Results}
\subsection{Topological defect accumulation in non-Hermitian systems with coupling non-reciprocity}
We consider the non-Hermitian SSH model \cite{su1979solitons, yao2018edge}, with the real-space Hamiltonian for an open chain of N unit cells given by 
\begin{equation}
\begin{aligned}
H = &\sum_{j=1}^{N-1} \left( t_1 |A_j\rangle\langle B_j| + t_1 |B_j\rangle\langle A_j| \right) \\
&+ \sum_{j=1}^{N-1} \left( (t_2 + \delta) |B_j\rangle\langle A_{j+1}| + (t_2 - \delta) |A_{j+1}\rangle\langle B_j| \right) \\
&+ \sum_{d} i\gamma_d |D_d\rangle\langle D_d|,
\end{aligned}
\end{equation}
where \(|A_j\rangle, |B_j\rangle\) are sublattice sites in unit cell j, \(t_1\) is intra-cell coupling, \(t_2 \pm \delta\) is non-reciprocal inter-cell coupling, and \(i\gamma_d\) is imaginary potential at defect site \(D_d\). The non-reciprocity \(\delta\) induces NHSE, while defects disrupt periodicity.

\begin{figure*}[htp!]
  \centering
    \includegraphics[width=0.85\textwidth]{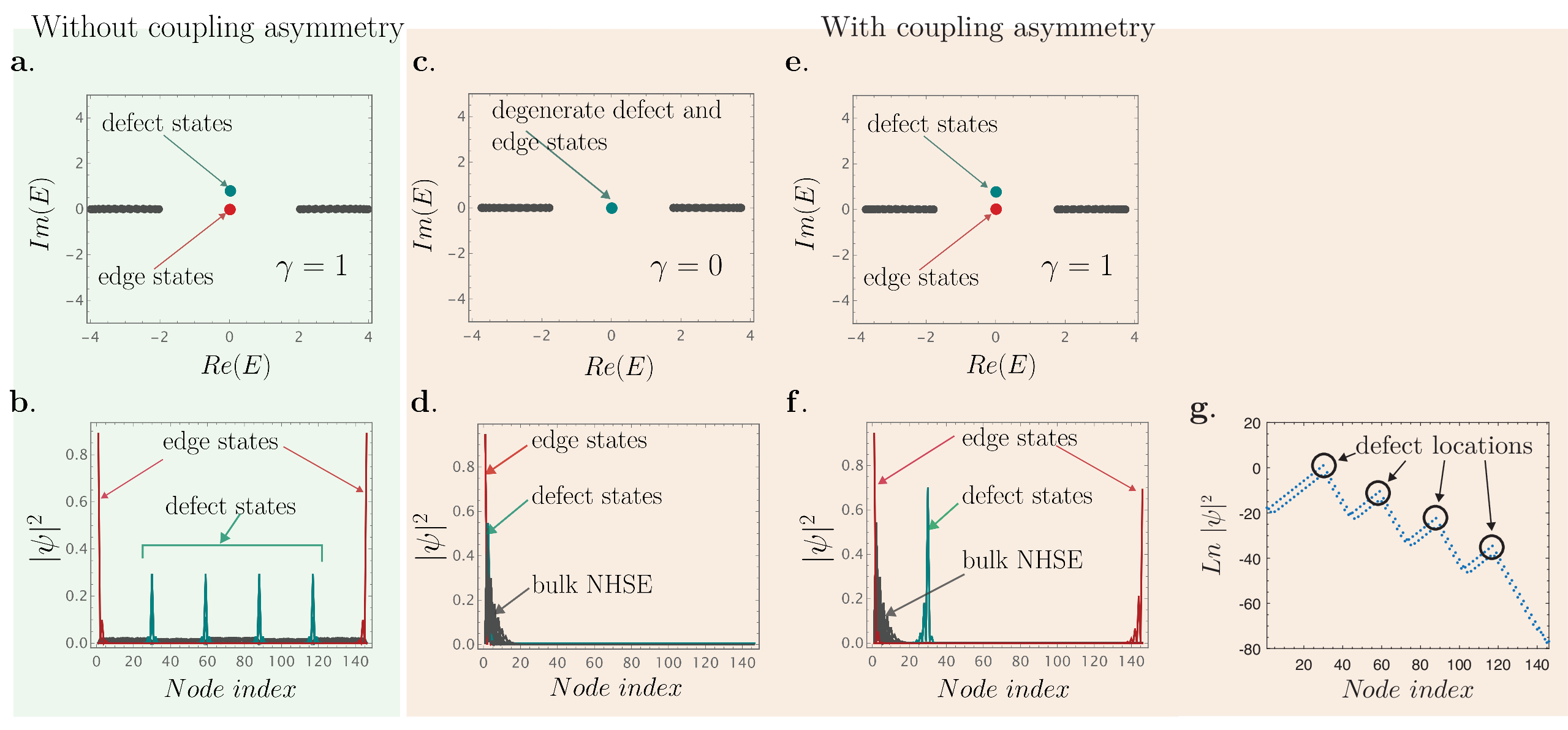}
  \caption{Origin and accumulation of \textcolor{blue}{NHSE-influenced} defect states in a non-Hermitian chain with four defect points and uniform loss $i\gamma$ at the defect nodes. (a) Complex eigenenergy and eigenstate spatial distribution in the Hermitian limit ($\gamma=1.0$, $\delta=0$) exhibiting four defect states with the energy of approximately $i\gamma$ in the middle of the energy gap and two edge states at zero energy. (b) Corresponding eigenstate spatial distributions localized at the respective defect points. (c-f) Eigenenergy and eigenstate distribution for defect states at different gain/loss $\gamma$ in the non-Hermitian chain with coupling asymmetry ($\delta=1.2$) exhibiting the accumulation of the defect states. In the absence of the onsite potential at the defect points ($\gamma=0$), the defect and edge states are degenerate at $E=0$ in the complex energy spectra (c) and localized at one of the edge nodes \textcolor{blue}{(due to NHSE)}, as shown in the corresponding eigenstate distribution (d). The midgap defect states are shifted towards a non-zero energy of approximately $E=i\gamma$ in the presence of onsite potential ($\gamma=1$) in (e), and the corresponding eigenstate distribution in (f) shows that the four defect states accumulate at only the defect point closest to the bulk NHSE localization edge. \textcolor{blue}{Edge states resist NHSE due to topological protection (zero energy pinned by chiral symmetry), while defect states at non-zero energy are influenced.} (g) shows a plot of the logarithm of the wavefunction amplitude, i.e., $\mathrm{\ln}\ |\psi|^2$ for one of the four degenerate states. (The other three defect states have similar $\mathrm{\ln}\ |\psi|^2$ profiles.) Common parameters: $t_1=1$, $t_2=3$.}
  \label{fig2}
\end{figure*}

To explain the origin of topological defect states and their accumulation in the presence of coupling asymmetry, we begin by examining a typical Hermitian SSH chain with the intra and inter-cell coupling strengths of $t_1$ and $t_2$, respectively (see Fig. \ref{fig1}a). The original SSH chain may host topological edge states, depending on the ratio of $t_1$ and $t_2$ and system's topology can be characterized by a  winding number. To study the influence of dissipative terms on the topological midgap states (i.e., defect states and edge states), we consider an open chain and insert an arbitrary number of defect sites with uniform loss factor of $i\gamma$ into the chain (see Fig. \ref{fig1}a). The emergence of topological defect states at the defect sites, arising from the breaking of the alternating sequence $t_1$ and $t_2$ coupling strengths, is determined by the symmetry of the underlying chain. 

Specifically, $n$ topological defect states appear in the eigenstate spectra (see Fig. \ref{fig1}a) if the pristine chain without the defects has a  completely real. To demonstrate the existence of defect states, we plot the complex eigenenergy and eigenvector spectra for a chain with $n=4$ defect points having a uniform loss factor of $i\gamma$ in Fig. \ref{fig2}a-b. The eigenenergy spectrum show four defect states with the same energy of approximately $i\gamma$ located in the midgap while there are two edge states at zero energy (see Fig. \ref{fig2}a). Spatially, the four defect states appear at the four defect sites while the two edge states are localized at the two boundaries (see Fig. \ref{fig2}b).

Next, we consider non-Hermitian systems, where the spatial distribution of the eigenstates  exhibit an exponential localization near a boundary under OBC known as  the NHSE. We study topological defect states in non-Hermitian systems by replacing the reciprocal inter-cell coupling $t_2$ (in Fig. \ref{fig1}a) with a directional unbalanced coupling $t_2 \pm \delta$ (see Fig. \ref{fig1}b). The corresponding Hamiltonian in $k$-space can be expressed using the Pauli matrices $\sigma_x$ and $\sigma_y$ as:
\begin{widetext}
\begin{equation}
H(k)= (t_1+t_2 \cos k_x - i \delta \sin k_x) \sigma_x +(t_2 \sin k_x + i \delta \cos k_x) \sigma_y.
\label{Eq1Ham}
\end{equation}
\end{widetext}
Here, $t_1$ and $t_2$ denote the intra- and inter-cell couplings, respectively, and $\delta$ represents the degree of non-reciprocity in the inter-unit cell couplings. This non-reciprocity may be realized in practice, for example, via the use of negative impedance converters at current inversion (INICs) in a TE realization \cite{rafi2024twisted,siu2023terminal,rafi2022interfacial,rafi2022unconventional,sahin2023impedance}.

We then insert an arbitrary number of defect nodes with the uniform loss factor of $i\gamma$ (indicated by the orange capsules in Fig. \ref{fig1}b) to investigate the interplay between non-Hermiticity and topological defect states in finite non-Hermitian SSH chains. These defect nodes disrupt the alternating series of $t_1$ and $(t_2 \pm \delta)$ couplings between neighboring nodes (refer to Fig. \ref{fig1}b).  The localization at the defect sites is induced by the presence of a gain / loss term $i\gamma$ at the defect nodes that differs from the zero onsite potential of the bulk nodes. To illustrate this phenomenon, we consider a specific scenario of four defect states interspersed over a non-Hermitian chain and plot its corresponding eigenenergy and eigenstate distribution at different gain/loss $\gamma$ values, as shown in Fig. \ref{fig2}c-f.

First, we examine the case of $\gamma=0$, where the onsite potentials at the bulk and defect nodes have the same value. In this situation, both edge states as well as all the defect states are located at $E=0$ in the complex energy spectra (Fig. \ref{fig2}c). All the edge and defect states, along with the NHSE bulk modes, are localized near the left or rightmost boundary of the chain, depending on the direction of the coupling asymmetry (Fig. \ref{fig2}d). Notably, all the bulk modes exhibit the same inverse decay length \cite{rafi2022critical} which depends on the coupling asymmetry, i.e., $r=\sqrt{\frac{t_2-\delta}{t_2+\delta}}$\textcolor{red}{ (derived in Supplementary by solving characteristic equation; GBZ is elliptic contour in complex \(\beta\)-plane with radius \(r\))}.
 
For a homogenous chain whose Hamiltonian is described by $H(\beta)$, $\beta\equiv\exp(i k)$,  the localization of its eigenstates can be characterized by the topological invariant $\eta$ defined as \cite{rafi2024twisted}
\begin{equation}
\eta = \frac{1}{2\pi} \oint_c  -i \partial_\beta \mathrm{ln}\  \left| H(\beta)\right|\ \mathrm{d}\beta.
\label{es3}
\end{equation}
where the integration contour is taken over the PBC eigenenergy ellipse on the complex energy plane. For the $\gamma=0$ case,  the sign of $\eta$ would determine whether the eigenstate localization occurs near the left or right edge of the chain, irrespective of the nature of the eigenstate, i.e., bulk, edge or defect. 

However, in the presence of a finite and uniform value of $\gamma$ at the defect sites,  the eigenenergies of the defect states are shifted to the vicinity of $E=i\gamma$ while the edge states remain at zero energy. (See the Supplementary Materials for approximate equations for the defect state eigenenergies.) The eigenenergies of the NHSE bulk skin modes are unaffected by the changes in the energy of the midgap states (refer to the eigenspectrum in Fig. \ref{fig2}e). Furthermore, the nearly degenerate defect states at energies close to $i\gamma$ are spatially localized at the left-most or right-most defect node in a manner consistent the NHSE localization edge of the bulk modes. For instance, in the case of negative value of $\eta$ (clockwise winding), the bulk modes are localized at the left edge of the chain. Likewise, all the defect states which were initially interspersed throughout the chain at their respective defect sites, coalesce onto the leftmost defect node closest to the left edge. Conversely, for positive values of $\eta$, the bulk modes are localized at the right edge of the chain, while all the defect states ``coalesce'' on the rightmost defect node. An illustration of this ``coalesced" defective mode localization is shown in Fig. \ref{fig2}f for the case of negative $\eta$. However, the two topological edge states remain at both the right and left extreme nodes of the open chain (Fig. \ref{fig2}f), resisting NHSE due to topological protection (zero energy pinned by chiral symmetry \(\Gamma H \Gamma^{-1} = -H\), \(\Gamma = \sigma_z\)), while defect states at non-zero energy are influenced.

Non-reciprocity biases propagation, 'pushing' defect modes toward the skin edge like unidirectional flow; direction controlled by sign of \(\delta\), flipping \(\eta\).

This accumulation of the defect states at the leftmost defect node may be explained intuitively by considering the spatial profile defect state wavefunctions. Figure \ref{fig2}g shows the logarithm of the wavefunction for one of the four nearly degenerate defect states with energies near $i\gamma$ shown in Fig. \ref{fig2}f. The logarithmic plot shows that the wavefunction amplitude first increases and then decreases in alternation along the length of the chain with local amplitude peaks localized around each defect. The amplitudes of these peaks decrease from left to right across successive defects because the slope of the linear increase, which corresponds to the logarithm of the larger of the two $|\beta|$ values in the SSH segment, is smaller than the slope of the linear decrease, which corresponds to the logarithm of the smaller $|\beta|$ value. This is explained in more detail in the Supplementary Materials, in which we also show that which of the two $\beta$ values has a larger slope is related to the NHSE localization direction of the pristine SSH chain without defects. 

The imaginary \(i\gamma\) shifts defect energies imaginarily, promoting degeneracy for coalescence. In contrast, complex onsite potentials $V+ i\gamma$
 shift defect energies to complex energy plane (i.e., $V + i\gamma$)
 and may slightly weaken accumulation due to hybridization, though clustering persists for small ($V$) (see Supplementary Fig. \ref{figS4} and Fig. \ref{figS5} for details).

Unlike dynamical edge burst \cite{xue2022non,xiao2024observation}, where loss amplifies at boundaries over time, our accumulation is static clustering at tunable defects.

\subsection{Energy-Dependent Defect State Accumulation in a Non-Hermitian Chain with Alternate Gain/Loss Modulation}
\begin{figure}[ht!]
  \centering
    \includegraphics[width=0.49\textwidth]{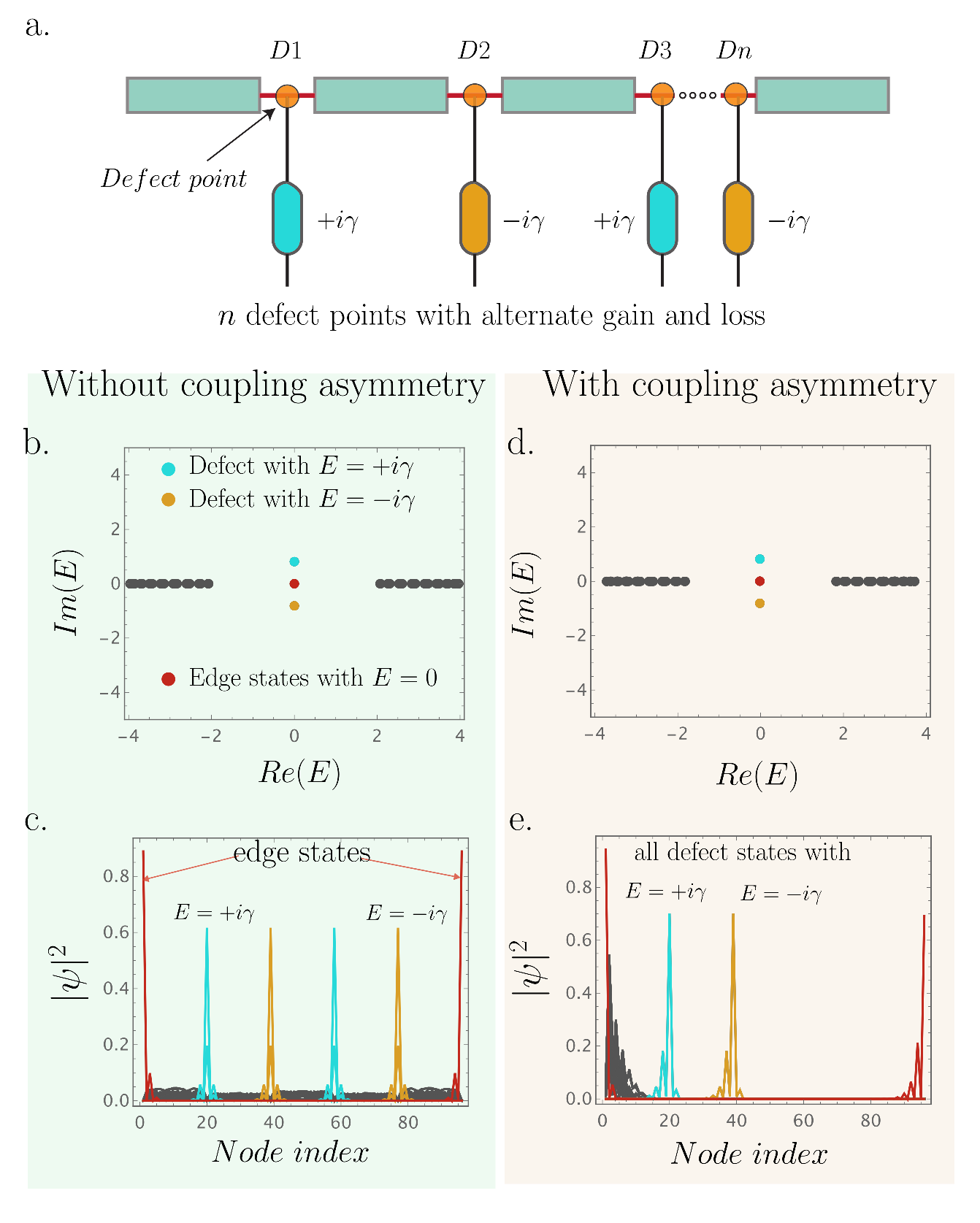}
  \caption{Defect state accumulation in a system containing $n=4$ defects alternating between gain and loss. (a) Schematic diagram of the modified mode. Loss ($i\gamma$) and gain ($-i\gamma$) are introduced alternatively at the odd- and even-numbered defect points, respectively. (b) Eigenenergy spectra of the modified chain in the Hermitian limit ($\delta=0$). $n/2$ degenerate defect states have eigenenergy at $-i\gamma$ while the remaining $n/2$ defect states have eigenenergy at $i\gamma$. The two edge states remain localized at zero energy. (c) Spatial distribution of the defective chain eigenstates in the Hermitian limit. The defect states are localized in the vicinity of their respective defect points. (d) Eigenenergy spectra of the modified chain in the non-Hermitian case ($\delta=1.2$). The eigenspectrum is similar to the Hermitian case, i.e., with $n/2$ degenerate defect states having eigenenergy at $-i\gamma$ while the remaining $n/2$ defect states have eigenenergy at $i\gamma$. (e) Spatial distribution of the defective chain eigenstates in the non-Hermitian case. Eigenstates with the same eigenenergies form clusters and are localized near the NHSE localization end. $n/2$ \textcolor{blue}{NHSE-influenced} defect states with the energy of approximately $i\gamma$ are localized at the first odd-numbered defect point, while the remaining $n/2$ \textcolor{blue}{NHSE-influenced} defect states with the energy of approximately $-i\gamma$ are accumulated at the first even-numbered defect point closest to the NHSE localization edge. Common parameters: $t_1=1$, $t_2=3$, and $\gamma=1$.}
  \label{fig3}
\end{figure}

Next, we investigate how the energy dependence and localization of the defect states are influenced by the spatial distribution of onsite gain / loss factors at the defect sites. We first modify our model by alternating the loss ($i\gamma$)  and gain ($-i\gamma$) on-site potentials at the odd and even-numbered defect nodes, respectively, as shown in Fig. \ref{fig3}a. We compare the resulting energy eigenspectra for the case of Hermitian chain (where the non-reciprocity factor $\delta = 0$) with that of a non-Hermitian chain (with finite $\delta$). For both cases, the energy eigenspectra appear identical, with $n/2$ degenerate defect states at eigenenergy of approximately $i\gamma$, while the other $n/2$ defect states at eigenenergy of approximately $-i\gamma$  (where $n$ is the total number of defect sites). However, the two topological edge states remain at zero energy (see Fig. \ref{fig3}b,d). Therefore, regardless of whether the chain is Hermitian or not, the defect states localized around defects with the same onsite potential (either $i\gamma$ or $-i\gamma$) form a cluster in the eigenenergy spectra near the corresponding onsite potential of the defects.

However, the spatial distribution of the defect states differs significantly in a non-Hermitian chain from that of its Hermitian counterpart (Fig. \ref{fig3}c,e). Specifically, in the Hermitian chain, the defect states are localized at their respective defect points (Fig. \ref{fig3}c). In contrast, for the non-Hermitian chain, the interplay between the defect states and the NHSE causes the defect eigenstates with the same energy to coalesce into a single cluster (Fig. \ref{fig3}e) on the corresponding defect site closest to the NHSE localization end. In our particular example, the $n/2$ (gainy) topological defect states with the energy of approximately $i\gamma$ are localized at the first odd-numbered defect point while the remaining $n/2$ (lossy) topological defect states with the energy of $-i\gamma$ are localized at the first even-numbered defect point closest to the NHSE localization edge (left boundary).

 \subsection{Effect of Dissimilar Gain/Loss Terms on the Localization of Defect States in Non-Hermitian Chains} 
  \begin{figure}[ht!]
  \centering
    \includegraphics[width=0.49\textwidth]{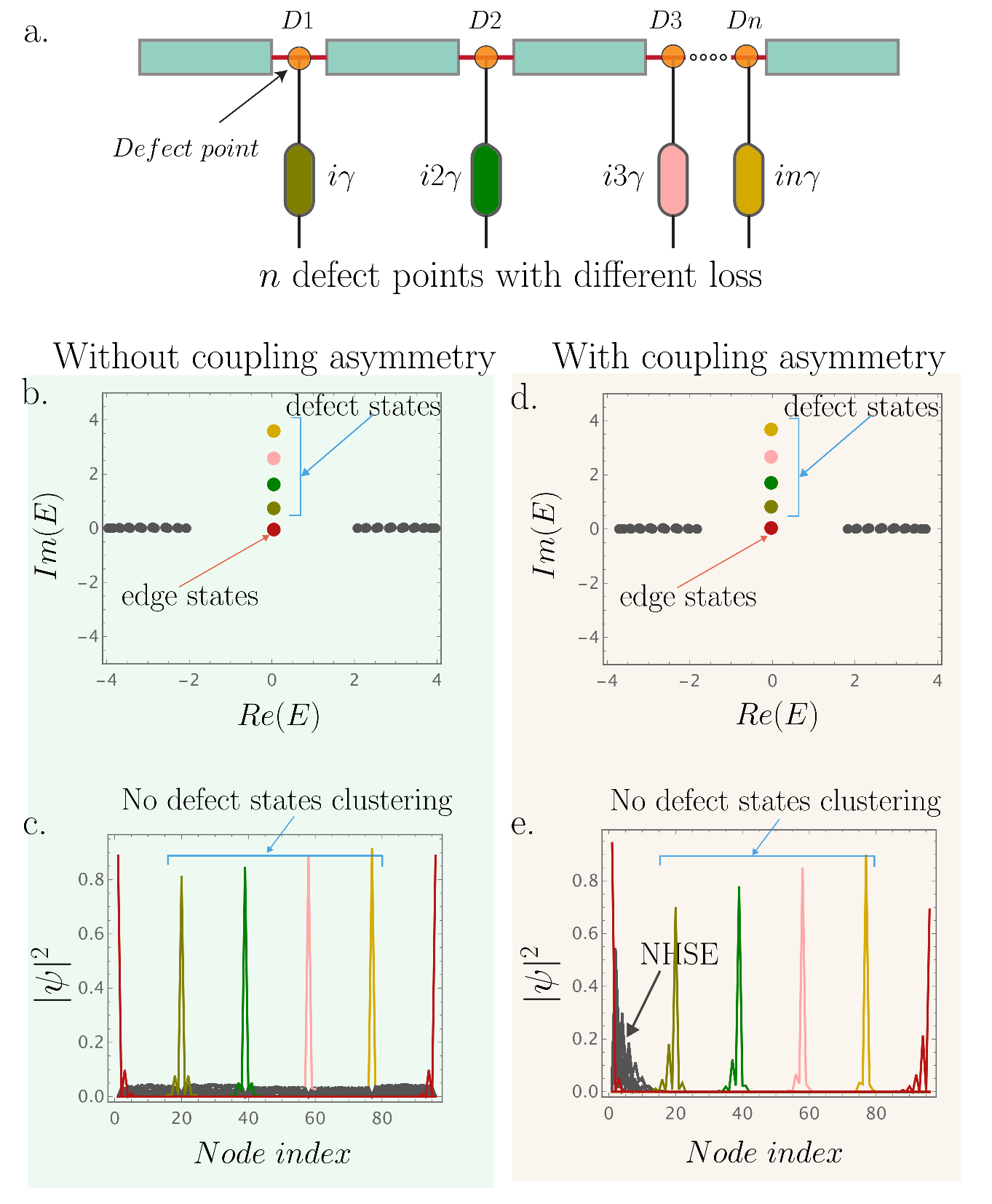}
 \caption{Defect state accumulation in a non-Hermitian chain with different gain/loss parameters at the defect nodes. (a) Schematic representation of a non-Hermitian chain containing $n$ defects with different loss parameters of $ij\gamma$ is introduced at the $j$th defect point ($j=1,2,3...n$). (b) Eigenenergy spectrum of the defective chain in the Hermitian case at $\delta=0$ in which the eigenenergies of the $n$ defect states are located at the $n$ distinct energy levels of approximately $E=ij\gamma$ in the complex energy plane. (c) The corresponding eigenstate profile shows that the $j$th defect state, which has an energy of approximately $E=ij\gamma$, is localized at the $j$th defect node, indicating the absence of defect state clustering. (d) The eigenenergy spectrum in the non-Hermitian scenario with $\delta=1.2$ and the presence of NHSE in the bulk modes of the chain shows $n$ midgap defect states at the $n$ defect points with dissimilar energy. (e) The spatial distribution of the eigenstates shows that the defect states do not cluster in space despite the presence of NHSE because the defect states are not degenerate in energy. All of the $n$ defect states have dissimilar energies and are separately localized at their respective defect points. \textcolor{blue}{This absence of clustering stems from energy-dependent \(\beta_{1,2}(E)\), prohibiting wavefunction mixing. A perturbation analysis confirms robustness: for degenerate cases, states remain coalesced under small disorder (\(\delta \gamma < 0.1 t_1\)); for non-degenerate, localization persists at individual sites (Supplementary Fig. S3).} Common parameters: $t_1=1$, $t_2=3$, and $\gamma=1$.}
  \label{fig4}
\end{figure}

In this section, we investigate the effect of a non-uniform gain/loss factor at the defect sites on the resulting spatial localization of the defect states. We introduce non-uniform loss term of $i (j \gamma)$ at the $j$th defect point where $j = 1, 2, ...n$ denotes the defect node index (i.e., right ascending loss profile shown in Fig. \ref{fig4}a). As in the previous cases of uniform and alternating gain/loss profiles (Figs. \ref{fig2}b, \ref{fig2}d and Figs. \ref{fig3}b, \ref{fig3}d, respectively), the resulting eigenspectra are identical for both cases of Hermitian and non-Hermitian chains (see Fig. \ref{fig4}b, \ref{fig4}d). Both eigenspectra show $n$ distinct defect state eigenenergies with values close to the corresponding loss potential values of $i (j \gamma)$. And as before, the energies of the two topological edge states are not influenced by the defect loss profile, and remain fixed at zero. One can surmise that when the loss potential at each defect site is distinct (i.e. given by $i (j \gamma)$), the energy degeneracy of the defect states is completely lifted and there is no longer any energy clustering. Interestingly, the spatial distribution of the defect states show a similar pattern in both the Hermitian and non-Hermitian chains (Figs. \ref{fig4}c, \ref{fig4}e). This is unlike the cases of uniform and alternating gain/loss profiles studied in the previous section (compare Figs. \ref{fig2}c, \ref{fig2}e with Figs. \ref{fig3}c, \ref{fig3}e, respectively). The NHSE localization is seen for the bulk modes for the non-Hermitian chain but not in the Hermitian one, as expected (compare Fig. \ref{fig4}c with Fig. \ref{fig4}e). However, surprisingly, no accumulation of defect states is observed for both the Hermitian and non-Hermitian chains. This is unlike the defective non-Hermitian chains studied in the previous sections. Thus, the spatial accumulation of defect states disappears when the energy degeneracy of the defect states are lifted. In this situation, the NHSE no longer influences the localization profile of defect states.

A corollary to the above results arises by considering a non-Hermitian chain with $n$ defect nodes/states which are divided into $m$ distinct clusters, i.e., where we set $m$ distinct values of the onsite potentials amongst the $n$ defect nodes. In such a scenario, the defect eigenstates  spatially coalesce into $m$ clusters. Each cluster consists of degenerate defect states (having the same eigenenergy) and localized at the defect node with the onsite potential corresponding to the cluster eigenenergy, and situated closest to the NHSE localization side. Conversely, if there are no degenerate defect states in the system, then the NHSE has no effect on the defect state distribution, since each defect state will localize separately at its own defect site without any spatial clustering.

This absence of clustering for non-degenerate states stems from the energy-dependent \(\beta_{1,2}(E)\), which prohibits wavefunction mixing. A perturbation analysis confirms this robustness, with degenerate states remaining coalesced under small random tolerances (2\% and 5\%) to parameters, as shown in Supplementary Fig. \ref{figS3}.

\subsection{Effect of Disorder on Defect Accumulation}

%\textcolor{red}{Disorder in \(\gamma\) lifts degeneracy, weakening accumulation; random positions scatter clusters, but direction follows NHSE Should we put this in the supplementary,,,, need to complete.}

\section{Significance  of Controlling Localization via Hermiticity and Onsite Defect Potentials}
The above results reveals the interplay of non-Hermiticity and onsite defect potentials and their impact  on the localization of defect states in topological systems. The ability to manipulate these factors opens up possibilities for precise control of the spatial distribution of defect states, which can be utilized for various applications \cite{rafiulislam2023twistedtopologybipolarnonhermitian,2022APS..MARN70013S,rafiulislam2021unconventionalnodevoltageaccumulation}.

Conversely, in Hermitian systems, defect states are localized at their respective defect nodes, maintaining their distinct spatial profiles irrespective of the system's overall topology or configuration. This predictable localization behavior is beneficial for designing stable and robust quantum devices where specific localization of states is required. The predictable nature of Hermitian systems ensures that defect states remain isolated, reducing unwanted interactions and preserving coherence in quantum information processing.

On the other hand, non-Hermitian systems possessing defect states that interact with the prevailing  defects, characterized by the NHSE, offer a unique advantage: the ability to cluster defect states at specific locations by tuning the gain/loss parameters. As described above in our results, this clustering effect causes defect states with degenerate eigenenergies to  accumulate near the NHSE localization edge. Such a clustering effect would be significant for applications in sensing and signal processing. We discuss some possible applications of defect states in non-Hermitian system:

\begin{itemize}
\item Topological sensors: The sensitivity of non-Hermitian systems to external perturbations can be exploited to design topological sensors with enhanced performance. By carefully tuning the gain/loss parameters, it is possible to localize defect states at predefined positions, in order to maximize interaction with external fields which are focused at those positions. This can lead to the development of sensors with high spatial resolution and sensitivity, capable of detecting minute changes in the environment.

\item	Quantum computing and information storage: In quantum computing, the ability to control the localization of defect states can be used to implement qubits and quantum gates with high precision \cite{wolfowicz2021quantum}. Hence, the  clustering of defect states at specific nodes can facilitate interactions between qubits, enabling efficient quantum gate operations. Moreover, the robustness of defect states in non-Hermitian systems can enhance the stability of quantum states, protecting them from decoherence and loss.

\item	Photonic devices and topolectrical circuits: The principles demonstrated in this study can be extended to design photonic and electronic devices with tailored properties. For instance, in photonic crystals, the localization of defect states can be controlled to create waveguides or resonators with specific characteristics \cite{khelif2003transmission}. Similarly, in electronic systems, the precise control over defect state localization can be used to engineer electronic band structures with desired non-Hermitian properties \cite{bai2018defect}, enhancing device performance.

\end{itemize}

\section{Conclusion}
In conclusion, we studied the effect of non-Hermiticity on the defect states localization in an open  chain with multiple topological defects. We demonstrated that the presence of non-Hermitian skin modes in the system leads to the to the spatial coalescence or clustering of defect states with degenerate eigenenergies onto the defect node with that degenerate eigenenergy which is located closest to the NHSE localization edge. In contrast, for the corresponding Hermitian chain, all the defect states are separately localized at their respective defect sites. We also analyzed and explained how the distribution of dissimilar gain/loss terms among defect points affects the energy clustering and spatial profile of the defect states.

Our results demonstrate that the NHSE can significantly influence the localization of topological defects in non-Hermitian system. The spatial clustering of degenerate defect states and their distinct localization behavior under the influence of gain/loss terms at the defect sites also open a new possibilities for the experimental detection and manipulation of topological defect states. 
 
Furthermore, our analysis and model are general and applicable to any synthetic platforms such as topolectrical circuits or photonic systems, which are experimentally accessible and may thus provide valuable insights into the properties of topological defects, and a path towards their potential applications.

\section*{Supplementary Materials}
\subsection{Defect state eigenenergy}
\begin{figure*}[htp!]
  \centering
    \includegraphics[width=0.75\textwidth]{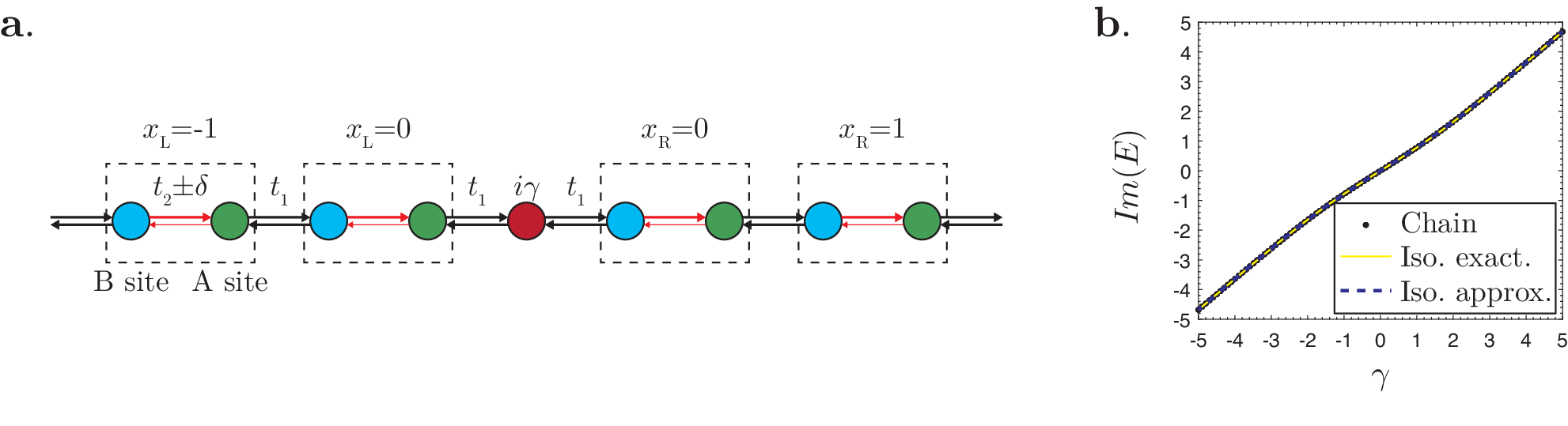}
  \caption{Isolated defect approximation. (a) Schematic diagram of an isolated defect sandwiched between semi-infinite non-Hermitian SSH chains with redefined unit cells (dotted boxes) each comprising a B sublattice site and the A sublattice site to its right. (b) Comparison of defect state eigenvalues for chain with four defects in Fig. \ref{fig2}g (`Chain'), the exact numerical solution of Eq. \eqref{eq:ConcCondE3} for the eigenenergy of an isolated defect (`\textcolor{red}{Isolated exact (Iso. exact.)}), and the linear approximation solution Eq. \eqref{eq:IsoLin} (`\textcolor{red}{Isolated approx. (Iso. approx.)}') at different values of $\gamma$. Common parameters: $t_1=1$, $t_2=3$, and $\delta=1.2$.}
  \label{figSuppSingDefect}
\end{figure*}

We first derive an approximate expression for the eigenenergy of a defect state with an on-site imaginary potential of $i\gamma$. To facilitate the analysis, we redefine the unit cell of the bulk SSH chain such that each unit cell contains a B sublattice site and the A sublattice site to its right, as shown in the dotted box in Fig. \ref{figSuppSingDefect}a. $t_1$ now becomes the inter-unit cell coupling and $t_2\pm\delta$ the intra-unit cell coupling, and the non-Bloch Hamiltonian $H'(\beta)$ with this new unit cell is given by 
\begin{equation}
  H'(\beta) = \begin{pmatrix} 0 & (t_2+\delta)+t_1/\beta  \\ (t_2-\delta)+t_1\beta  & 0 \end{pmatrix}  \label{eq:HpBeta}
\end{equation}
in the B-A sublattice basis. 

The eigenstate wavefunction $\psi(x)$ within the bulk of a SSH chain for a given eigenenergy $E$ has the general form
\begin{equation}
  \psi(x) = c_1 \beta_1^x \begin{pmatrix} 1 \\ \chi_1 \end{pmatrix} + c_2 \beta_2^x \begin{pmatrix}1 \\ \chi_2\end{pmatrix}
\end{equation}
\textcolor{red}{where \(\beta_{1,2}\) are obtained by solving \(\det(E - H'(\beta)) = 0\), with \(|\beta_2| \geq |\beta_1|\), and \(\chi_j\) are the corresponding eigenspinors.}
where $\beta_{(1,2)}$ are the two solutions of $\beta$ for $|E-H'(\beta)|=0$ with $|\beta_2|\geq|\beta_1|$, and $c_j$ is the weight of the $\beta_j$ component in the wavefunction and $(1, \chi_j)^{\mathrm{T}}$ its corresponding eigenspinor.   

The near linearity and local peaking of the logarithm-scale density distributions in the vicinities of each of the defects in Fig. \ref{fig2}g imply that only the $\beta_2$ component ($\beta_1$) has a significant contribution to the SSH wavefunction to the left (right) of each defect. A reasonable first approximation is therefore to set the weight of the $\beta_1$ component to the left of each defect to exactly 0 and that of the $\beta_2$ component to its right to 0. This is equivalent to assuming that the $\beta_1$ component from the left neighboring defect and $\beta_2$ component from the right neighboring defect have decayed to 0 over the length of the SSH segments between the defects. Each defect is therefore effectively isolated from its neighbors and behaves as if it is sandwiched between two semi-infinite SSH segments. 

We consider one such isolated defect and solve for its eigenenergy. Writing the SSH wavefunction to the left of the defect as $\psi_{\mathrm{L}}(x_{\mathrm{L}}) = \sum_{j=1}^2 c_{\mathrm{L}, j} \beta_j^{x_{\mathrm{L}}} (1, \chi_j)^{\mathrm{T}}$ and that to the right as $\psi_{\mathrm{R}}(x_{\mathrm{R}}) = \sum_{j=1}^2 c_{\mathrm{R}, j} \beta_j^{x_{\mathrm{R}}} (1, \chi_j)^{\mathrm{T}}$ where we set $x_{\mathrm{L}}=0$ at the unit cell to the immediate left of the defect and $x_{\mathrm{R}}=0$ at that to the immediate right of the defect, the Schroedinger equation at the defect and the two sites abutting it lead to the three boundary conditions
\begin{align}
  & \langle\mathrm{B}|\psi_{\mathrm{L}}(1)= \psi_{\mathrm{D}} \label{eq:BC1} \\
  &t_1 \left( \langle A|\psi_{\mathrm{L}}(0) + \langle B|\psi_{\mathrm{L}}(0) \right) + (i\gamma-E)\psi_{\mathrm{D}} = 0 \label{eq:BC2}\\
  &\langle \mathrm{A}|\psi_{\mathrm{R}}(-1)= \psi_{\mathrm{D}} \label{eq:BC3}.
\end{align}  
where $\psi_D$ is the wavefunction amplitude at the defect node. 

The isolated defect approximation corresponds to setting $c_{\mathrm{L},1}=0$ and $c_{\mathrm{R},2}=0$. Without loss of generality, we further let $c_{\mathrm{L},2}=1$. Substituting these values of $c_{\mathrm{L},1}$, $c_{\mathrm{R},2}$, and $c_{\mathrm{L},2}$ into Eq. \eqref{eq:BC2} and \eqref{eq:BC3} results in a system of two linear equations that can be solved to yield $c_{\mathrm{R},1}$ and $\psi_{\mathrm{D}}$. Requiring Eq. \eqref{eq:BC1} to be satisfied for the resultant $c_{\mathrm{L},1}$, $c_{\mathrm{R},2}$, $c_{\mathrm{L},2}$, $c_{\mathrm{R},1}$, and $\psi_{\mathrm{D}}$ gives the consistency condition 
\begin{equation}
  (i\gamma-E) + t_1 \left( \frac{\chi_2}{\beta_2} + \frac{\beta_1}{\chi_1} \right) = 0. \label{eq:ConcCond}
\end{equation}    
Recall that the $\chi_j$s and $\beta_j$s are themselves functions of $E$. Substituting their explicit forms in terms of $E$, $t_1$, $t_2$, and $\delta$ into Eq. \eqref{eq:ConcCond} and simplifying results in a third-order equation in $E$,
\begin{equation}
  E^3 + \left(2\delta^2 +\gamma^2 - 2(t_1^2+t_2^2)\right)E - 2i\gamma\left(\delta^2+t_1^2-t_2^2\right) = 0. \label{eq:ConcCondE3} 
\end{equation}
Two of the solutions for $E$ are complex and one is purely imaginary. The purely imaginary solution for $E$ is the eigenenergy of the defect state under the isolated defect approximation. Unfortunately, Eq. \eqref{eq:ConcCondE3} does not have a simple analytic solution. To remedy this, we note from the numerical results in the main manuscript that the eigenenergies of the defect states are approximately equal to $i\gamma$. We therefore write $E=i(\gamma+\delta E)$, retain only the linear terms in $\delta E$ in Eq. \eqref{eq:ConcCond}\textcolor{red}{ (substituting \(E = i\gamma + i\delta E\) into Eq. \eqref{eq:ConcCond} and expanding to first order in \(\delta E\))}, and solve for $\delta E$ in the resultant expression to obtain 
\begin{widetext}
\begin{align}
  \delta E = & t_1^2 \left( \delta^2 - \gamma^2 + t_1^2 - t_2^2 
  + \sqrt{ (\delta^2 - (\gamma - i t_1)^2 - t_2^2) 
  (\delta^2 - (\gamma + i t_1)^2 - t_2^2) } \right) \notag \\
  & \times \left[ 
  \frac{ \gamma t_1^2 - 
  2 \gamma t_1^2 
  \left( \delta^2 - (\gamma + i t_1)^2 - t_2^2 \right)^{-1} 
  }{ - \delta^2 + \gamma^2 t_1^2 t_2^2 
  + \sqrt{ \delta^4 + (\gamma^2 + t_1^2)^2 
  + 2 (\gamma - t_1)(\gamma + t_1) t_2^2 
  + t_2^4 - 2 \delta^2 (\gamma^2 - t_1^2 + t_2^2) } 
  } \right]^{-1} \notag \\
  & \times \frac{1}{\gamma} t_1^2 
  \left( \delta^2 - \gamma^2 + t_1^2 - t_2^2 
  + \sqrt{ \delta^4 + (\gamma^2 + t_1^2)^2 
  + 2 (\gamma - t_1)(\gamma + t_1) t_2^2 
  + t_2^4 - 2 \delta^2 (\gamma^2 - t_1^2 + t_2^2) } 
  \right). \label{eq:IsoLin}
\end{align}
\end{widetext}

Figure \ref{figSuppSingDefect}b shows a comparison of the numerically obtained defect state eigenenergies of the chain with four defects in Fig. \ref{fig2}g, the exact isolated defect approximation eigenenergy obtained via the numerical solution of Eq. \eqref{eq:ConcCondE3}, and the linear approximation isolated defect approximation eigenenergy Eq. \eqref{eq:IsoLin} at different values of $\gamma$. The curves are visually indistinguishable from one another at the scale of the plot. The close match between them demonstrates the validity of the isolated defect approximation.

The parameter \(E_0 = 0.784\) is the isolated defect approximation eigenenergy obtained numerically from Eq. \eqref{eq:ConcCondE3} for the specific parameters \(t_1=1\), \(t_2=3\), \(\delta=1.2\), \(\gamma=1\). It serves as the base value around which small deviations occur in the multi-defect case.

\subsection{Decay Length Derivation}

The decay length, defined as \( r = \sqrt{\frac{t_2 - \delta}{t_2 + \delta}} \), characterizes the spatial localization of the skin modes in a non-Hermitian system under the non-Bloch band theory. This expression is derived by analyzing the characteristic equation for the non-Bloch wave number \(\beta\), where the radius of the generalized Brillouin zone (GBZ) determines the extent of the skin localization. Below, we provide a detailed derivation of this decay length.

Consider a one-dimensional non-Hermitian tight-binding model, such as the Hatano-Nelson model or a similar system with asymmetric hopping. The Hamiltonian may include terms with non-reciprocal hopping amplitudes, leading to the non-Hermitian skin effect, where eigenstates localize at the boundaries. The non-Bloch wave number \(\beta\) is a complex quantity that describes the spatial profile of these eigenstates, and the decay length is related to the magnitude of \(\beta\), i.e., \( |\beta| = r \), which defines the GBZ radius.

The characteristic equation for the system is typically obtained from the eigenvalue problem of the non-Hermitian Hamiltonian. For a simplified model, consider a lattice with asymmetric hopping amplitudes \( t_1 = t_2 + \delta \) and \( t_2 - \delta \), where \( t_2 \) is the average hopping strength and \( \delta \) accounts for the non-Hermiticity (asymmetry in hopping). The characteristic equation for the wave number \(\beta\) in such a system can be written as:

\[
E = (t_2 + \delta) \beta + (t_2 - \delta) \beta^{-1},
\]

where \( E \) is the energy eigenvalue. To find the decay length, we focus on the spatial behavior of the wave function, which is determined by the magnitude of \(\beta\). The non-Bloch wave number \(\beta\) lies on the GBZ, defined by the condition \( |\beta| = r \), where \( r \) is the radius of the GBZ in the complex plane.

To derive \( r \), we solve for \(\beta\) such that \( |\beta| = r \). Multiplying the characteristic equation by \(\beta\), we obtain a quadratic equation in \(\beta\):

\[
(t_2 - \delta) + E \beta + (t_2 + \delta) \beta^2 = 0.
\]

The solutions to this quadratic equation are:

\[
\beta = \frac{-E \pm \sqrt{E^2 - 4 (t_2 + \delta)(t_2 - \delta)}}{2 (t_2 + \delta)}.
\]

The GBZ is determined by the condition that the magnitudes of the two solutions, \( \beta_+ \) and \( \beta_- \), are equal, i.e., \( |\beta_+| = |\beta_-| = r \). This ensures that the eigenstates decay with the same characteristic length on both sides of the system (under open boundary conditions). The product of the roots of the quadratic equation gives:

\[
\beta_+ \beta_- = \frac{t_2 - \delta}{t_2 + \delta}.
\]

Since \( |\beta_+| = |\beta_-| = r \), we have \( |\beta_+ \beta_-| = r^2 \). Taking the absolute value of the product:

\[
r^2 = \left| \frac{t_2 - \delta}{t_2 + \delta} \right|.
\]

Assuming \( t_2 \) and \(\delta\) are real and \( t_2 > |\delta| \) (ensuring positive arguments), we obtain:

\[
r = \sqrt{\frac{t_2 - \delta}{t_2 + \delta}}.
\]

This \( r \) represents the decay length of the skin modes, as it determines the exponential decay rate of the wave function, \( \psi_n \sim r^{|n|} \), where \( n \) is the lattice site index. The GBZ radius \( r \) quantifies the localization strength of the non-Hermitian skin effect, with \( r < 1 \) indicating localization toward one boundary and \( r > 1 \) indicating localization toward the opposite boundary.

In summary, the decay length \( r = \sqrt{\frac{t_2 - \delta}{t_2 + \delta}} \) is obtained by solving the characteristic equation for the non-Bloch wave number \(\beta\), with the GBZ radius determining the skin localization. This derivation assumes a simplified non-Hermitian tight-binding model, but the framework can be extended to more complex systems by analyzing the corresponding characteristic equations.

\subsection{Eigenstate coalescence}
Although the isolated defect approximation provides a good approximation for the  the chain defect state eigenenergies, it cannot explain why the eigenstates coalesce towards one particular boundary of the chain via the decay of the density peaks at successive defects away from the boundary. In the isolated defect approximation, it is assumed that the SSH wavefunction on the left of each defect contains only a $\beta_2$ component while that on its right contains only a $\beta_1$ component. This cannot be the case in a chain containing multiple defects because requiring the SSH segment to the right of a defect to contain only a $\beta_1$ component is inconsistent with requiring that the same SSH segment, which is on the left of the right neighboring defect, contains only a $\beta_2$ component. In the actual chain, the SSH segments between the defects and boundaries have finite weights of both the $\beta_1$ and $\beta_2$ components, and the eigenenergies of the defect states deviate slightly from that obtained using the isolated defect approximation. For instance, the deviations from the isolated defect approximation eigenenergy of 0.784 for the chain shown in Fig. \ref{fig2}g are on the order of $2\times10^{-11}$.

The parameter \(E_0 = 0.784\) originates from the numerical solution of the isolated defect eigenenergy Eq. \eqref{eq:ConcCondE3} for parameters \(t_1=1\), \(t_2=3\), \(\delta=1.2\), \(\gamma=1\). It represents the base eigenenergy for a single defect, with small perturbations in multi-defect systems.

\begin{figure*}[htp!]
  \centering
    \includegraphics[width=0.75\textwidth]{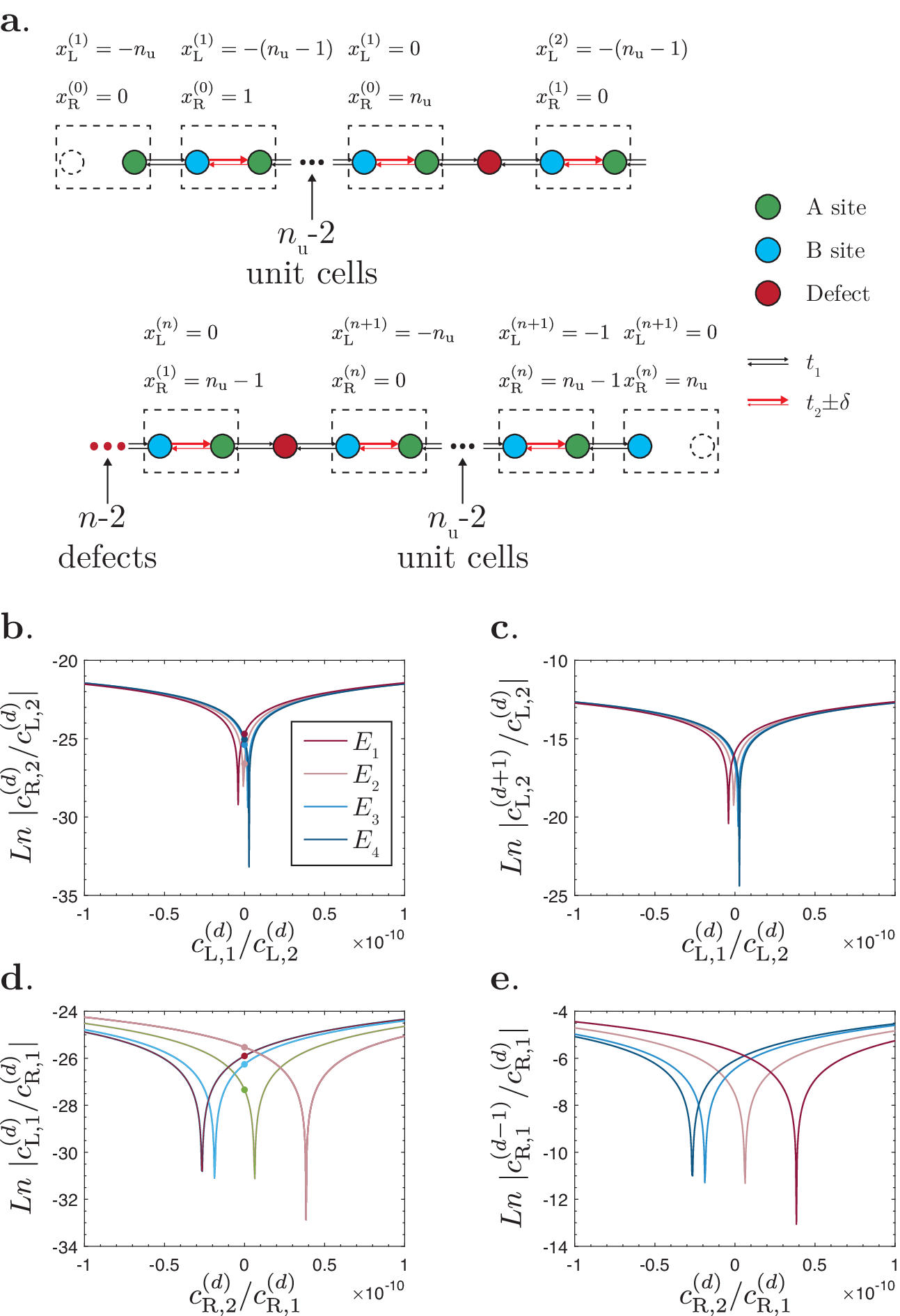}
  \caption{Chain with multiple defects. (a) Schematic representation of a chain with $n$ defects and $n_{\mathrm{u}}$ complete B-A unit cells in the SSH chain between successive defects or between a boundary and its immediate neighboring defect, and illustration of the $x^{(d)}_{(\mathrm{L},\mathrm{R})}$ coordinate system. (b) Resultant $\beta_2$ weight on the right of defect $d$, $c^{(d)}_{\mathrm{R}, 2}$ as function of the $\beta_1$ weight on the left of the defect, $c^{(d)}_{\mathrm{L},1}$ per unit input of $\beta_2$ on the left of the defect $c^{(d)}_{\mathrm{L},2}$ at the four defect state eigenenergies $E_1$--$E_4$ of the chain in Fig. \ref{fig2}g. The circles denote the values of $|c^{(1)}_{\mathrm{R}, 2}/c^{(1)}_{\mathrm{L},2}|$ and $c^{(1)}_{\mathrm{L}, 1}/c^{(1)}_{\mathrm{L},2}$ at the left-most defect $d=1$ in the chain. (c) Corresponding resultant $\beta_2$ weights on the left of defect $d+1$, $c^{(d+1)}_{\mathrm{L}, 2} = \beta_2^{n_{\mathrm{u}}-1}c^{(d)}_{\mathrm{R},2}$ per $c^{(d)}_{\mathrm{L},2}$. (d) Resultant $\beta_1$ weight on the left of defect $d$, $c^{(d)}_{\mathrm{L}, 1}$ as function of the $\beta_2$ weight on the right of the defect, $c^{(d)}_{\mathrm{R},2}$ per unit input of $\beta_1$ on the right of the defect $c^{(d)}_{\mathrm{R},1}$ at the four defect state eigenenergies $E_1$--$E_4$ of the chain in Fig. \ref{fig2}g. The circles denote the values of $|c^{(d)}_{\mathrm{L}, 1}/c^{(d)}_{\mathrm{R},1}|$ and $c^{(n)}_{\mathrm{R}, 2}/c^{(n)}_{\mathrm{R},1}$ at the right-most $d=n$ defect in the chain. (e) Corresponding $\beta_1$ weights on right of defect $d-1$, $c^{(d-1)}_{\mathrm{R}}=\beta_1^{-(n_{\mathrm{u}}-1)}c^{(d)}_{\mathrm{L}, 1}$. Common parameters: $t_1=1$, $t_2=3$, $\delta=1.2$, $\gamma=1$, and $n_{\mathrm{u}}=14$.  }
  \label{figSuppSys}
\end{figure*}

For the convenience of the subsequent discussion, we extend the notation for the wavefunction and coordinate system, as shown in Fig. \ref{figSuppSys}a. We introduce the superscript index $(d)$ to $x_{(\mathrm{L},\mathrm{R})}$ and $c_{(\mathrm{L},\mathrm{R}), j}$ such  that $x^{(d)}_{\mathrm{L}}$ ($x^{(d)}_{\mathrm{R}}$) is the unit cell index to the left (right) of defect $d$ with $x^{(d)}_{\mathrm{L}}=0$ ($x^{(d)}_{\mathrm{R}}=0$) being the unit cell to the immediate left (right) of the $d$th defect, and $c^{(d)}_{\mathrm{(L,R)},j}$ is the weight of the $\beta_j$ component in the $x^{(d)}_{\mathrm{(L,R)},j}$ coordinate system. We label the left and right boundaries comprising incomplete B-A unit cells with a missing A site at the left boundary and B site at the right boundary as the $d=0$th and $d=(n+1)$th defects, respectively, and set $x^{(0)}_{\mathrm{R}}=0$ at the left incomplete unit cell and $x^{(n+1)}_{\mathrm{L}}=0$ at the right incomplete unit cell (see Fig. \ref{figSuppSys}a) for convenience. We consider a chain similar to the one in Fig. \ref{fig2}g in which there are $n_{\mathrm{u}}$ complete unit cells in every SSH segment between two defects. In this case, because the right side of defect $d$ is the left side of its right neighbor defect $d+1$, $x^{(d)}_{\mathrm{R}}$ and $x^{(d+1)}_{\mathrm{L}}$ are related via $x^{(d)}_{\mathrm{R}} = n_{\mathrm{u}}-1+x^{(d+1)}_{\mathrm{L}}$. % whereas the consistency requirement $c^{(d)}_{\mathrm{R},j}\beta^{x^{(d)_{\mathrm{R}}}}_j = c^{(d+1)}_{\mathrm{L},j}\beta^{x^{(d+1)_{\mathrm{L}}}}_j$ implies that  $= c^{(d+1)}_{\mathrm{L},j} = c^{(d)}_{\mathrm{R},j}\beta_j^{n_{\mathrm{u}}-1}$.

The multiple defects in a chain interact with one another in the following way:
The requirement that the wavefunction vanishes at the site to the immediate left of the left chain boundary can be written as $\langle \mathrm{B}|\psi^{(0)}_{\mathrm{R}}(x^{(0)}_{\mathrm{R}}=0)=0$, which gives $c^{(0)}_{\mathrm{R},1}=-c^{(0)}_{\mathrm{R},2}$. The $\beta_1$ and $\beta_2$ components propagate from the left chain boundary to the left of the leftmost defect, giving rise to $c^{(1)}_{\mathrm{L},1}/c^{(1)}_{\mathrm{L},2} = (\beta_1/\beta_2)^{n_{\mathrm{u}}} (c^{(0)}_{\mathrm{R},1}/c^{(0)}_{\mathrm{R},2}) = -(\beta_1/\beta_2)^{n_{\mathrm{u}}}$. % really is nu here. cos 0 is the half unit cell. 

% whoops. it turns out that actually my desktop and laptop give slightly different answers and plots sia.

For the exemplary chain in Fig. \ref{fig2}g, $|\beta_1|\approx 0.218$, $|\beta_2|\approx 1.97$, and $|c^{(1)}_{\mathrm{L},1}/c^{(1)}_{\mathrm{L},2}|$ is on the order of $4.22\times 10^{-14}$ across all the defect states. The very small magnitude of $c^{(1)}_{\mathrm{L},1}$ relative to $c^{(1)}_{\mathrm{L},2}$ explains the close match between the approximate eigenenergy obtained using the isolated defect eigenenergy approximation and the exact eigenenergies of the defect states. Given $c^{(1)}_{\mathrm{L},1}$ and $c^{(1)}_{\mathrm{L},2}$ at a particular $E$ on the left of the first defect, the defect boundary conditions Eq. \eqref{eq:BC1} to \eqref{eq:BC3} constitute a system of three linear equations that can be solved for the defect wavefunction amplitude $\psi^{(1)}_{\mathrm{D}}$ as well as $c^{(1)}_{\mathrm{R},1}$ and $c^{(1)}_{\mathrm{R},2}$ to the right of the defect. (We omit their explicit analytical expressions here because they are very complicated and not very illuminating. ) For a given value of $c^{(d)}_{\mathrm{L},1}/c^{(d)}_{\mathrm{L},2}$, the resultant $c^{(d)}_{\mathrm{R},j}$s on the right of a defect are proportional to $c^{(d)}_{\mathrm{L},2}$ on the left. Figure \ref{figSuppSys}b shows the output $c^{(d)}_{\mathrm{R},2}$ per unit input $c^{(d)}_{\mathrm{L},2}$ as a function of $c^{(d)}_{\mathrm{L},1}/c^{(d)}_{\mathrm{L},2}$ for the four defect state eigenenergies $E_j=E_0+(-0.269, -0.0363, 0.135, 0.175)\times 10^{-10}$ of the chain in Fig. \ref{fig2}g  where $E_0=0.784$ is the isolated defect approximation eigenenergy. The values of $c^{(1)}_{\mathrm{L},1}/c^{(1)}_{\mathrm{L},2}$ at the leftmost defect are indicated by the filled circles. Note that although $c^{(d)}_{\mathrm{L},1}/c^{(d)}_{\mathrm{L},2}$ may vary across different defects in a chain, Fig. \ref{figSuppSys}b shows that $\mathrm{ln}\ |c^{(d)}_{\mathrm{R},2}/c^{(d)}_{\mathrm{L},2}|$ converges to a constant value at larger values of $|c^{(d)}_{\mathrm{L},1}/c^{(d)}_{\mathrm{L},2}|$. The fact that $|c^{(d)}_{\mathrm{R},2}/c^{(d)}_{\mathrm{L},2}|$  approaches 0 at small values of $|c^{(d)}_{\mathrm{L},1}/c^{(d)}_{\mathrm{L},2}|$ is expected from the very small deviations of the $E_j$s from $E_0$, at which $c^{(d)}_{\mathrm{R},2}$ is exactly 0 when $c^{(d)}_{\mathrm{L},1}$ is 0 by definition. 

Despite the very small values of the output $c^{(1)}_{\mathrm{R}, 2}$ at the right of the first defect relative to that of the input $c^{(1)}_{\mathrm{R}, 1}$ on its left, the $\beta_2$ component in the second SSH segment is scaled up by $\beta_2^{n_{\mathrm{u}}-1}$ as it propagates to the right across the SSH segment to the second defect. This scaling results in a $\beta_2$ weight of $c^{(2)}_{\mathrm{L},2} =c^{(1)}_{\mathrm{R},2}\beta_2^{n_{\mathrm{u}}-1}$ at the left of the second defect. The $\beta_2$ component at the left of the second defect in turn gets transmitted through the defect and propagates through the third SSH segment to the third defect, and the process of scaling and transmission repeats iteratively for successive defects. At each iteration, the $\beta_2$ input at the left of the $(d+1)$th defect is scaled relative to that at the $d$th defect by $c^{(d+1)}_{\mathrm{L},2}/c^{(d)}_{\mathrm{L},2}=\beta_2^{(n_{\mathrm{u}}-1)}c^{(d)}_{\mathrm{R},2}/c^{(d)}_{\mathrm{L},2}$. Figure \ref{figSuppSys}(c) shows that $|c^{(d+1)}_{\mathrm{L},2}/c^{(d)}_{\mathrm{L},2}| \ll 1$. This implies that the amount of incoming excitation at the left of each defect tends to become smaller at successive defects towards the right, which drives the eigenstate wavefunction amplitude down from left to right. (Nb. although it might appear that $|c^{(d+1)}_{\mathrm{L},2}/c^{(d)}_{\mathrm{L},2}|$ can made to exceed 1 by increasing $n_{\mathrm{u}}$ so that the $\beta_2$ component is amplified more as it travels through the SSH segment, in practice this does not occur because the eigenenergies are then driven closer to $E_0$ and $|c^{(d)}_{\mathrm{R},2}/c^{(d)}_{\mathrm{L},2}|$ becomes closer to 0 to compensate. )

A similar analysis can be performed starting from the right boundary. The boundary condition that $\langle\mathrm{A}|\psi^{(n+1)}_{\mathrm{R}} = 0$  leads to $c^{(n)}_{\mathrm{R},2}/c^{(n)}_{\mathrm{R},1} = -(\chi_1/\chi_2) (\beta_2/\beta_1)^{-{n_\mathrm{u}}}$ at the right of the rightmost $d=n$ defect, which is on the order of $-4.22\times 10^{-13}$. The incoming $\beta_1$ component on the right of the rightmost defect is transmitted through the defect, scaled up by $\beta_1^{-(n_{\mathrm{u}}-1)}$ as it travels towards its left neighbor the $(n-1)$th defect, following which it undergoes repeated transmission and scaling as it propagates further towards the left across successive defects. For this analysis, we take the component weights on the right of each defect $c^{(d)}_{\mathrm{R},j}$ to be given and solve the boundary equations Eq. \eqref{eq:BC1} to \eqref{eq:BC3} for $\psi_{\mathrm{D}}$ and $c^{(d)}_{\mathrm{L},j}$.  Figure \ref{figSuppSys}d shows the weight of the $\beta_1$ component transmitted into the SSH segment to the left of the $d$th defect $c^{(d)}_{\mathrm{L},1}$ per unit input of the $\beta_1$ component $c^{(d)}_{\mathrm{R},1}$ in the SSH segment to the right of the defect as a function of $c^{(d)}_{\mathrm{R},2}/c^{(d)}_{\mathrm{R},1}$ with the values of $c^{(d)}_{\mathrm{R},2}/c^{(d)}_{\mathrm{R},1}$ at the rightmost defect indicated by the solid circles. Comparing Fig. \ref{figSuppSys}b and \ref{figSuppSys}d, it can be seen that the logarithms of the outgoing wavefunction component weights per unit incoming wavefunction component at the left-most and right-most defects have approximately the same magnitudes of $-26\pm 1$. However, the larger magnitude of $|\mathrm{ln}\ |\beta_1||=1.52$ compared to $|\mathrm{ln}\ |\beta_2||=0.676$ implies that the $\beta_1$ component is more significantly amplified by $\beta_1^{-(n_{\mathrm{u}-1})}$ as it propagates across the SSH segment from a defect to its left neighboring defect compared to the amplification of $\beta_2^{n_{\mathrm{u}}-1}$ that the $\beta_2$ component undergoes as it propagates from a defect to its right neighbor. Figure \ref{figSuppSys}e shows that as a result, the amount that the incoming $\beta_1$ component at the right of each defect is scaled by across successive defects, $c^{(d-1)}_{\mathrm{R},1}/c^{(d)}_{\mathrm{R},1}$, has a much larger magnitude compared to that of $c^{(d+1)}_{\mathrm{L},2}/c^{(d)}_{\mathrm{L},2}$ shown in Fig. \ref{figSuppSys}c. Although $|c^{(d-1)}_{\mathrm{R},1}/c^{(d)}_{\mathrm{R},1}| < 1$ has a tendency to drive the eigenstate wavefunction amplitude down from the right to the left, this effect is in competition with the tendency of  $|c^{(d+1)}_{\mathrm{L},2}/c^{(d)}_{\mathrm{L},2}| < 1$ to drive the eigenstate wavefunction amplitude down from the left to the right. In this competition, the latter prevails because of its larger wavefunction amplitude drop across successive defects. The wavefunction amplitude therefore decreases from left to right, and the defect states are localized at the leftmost defect. 

Having shown that whether the defect state wavefunction amplitudes decrease towards the left or right in a chain with regularly spaced defects is predominantly determined by which of $|\mathrm{ln}\ |\beta_j||$ has a larger magnitude, we now briefly discuss the values of $|\mathrm{ln}\ |\beta_j||$. If $\mathrm{ln}\ |\beta_2|$ and $\mathrm{ln}\ |\beta_1|$ are both positive, the wavefunction amplitude unambiguously grows from left to right and the defect states are localized towards the right. By the definition that $|\beta_2| \geq |\beta_1|$, $|\mathrm{ln}\ |\beta_2||> |\mathrm{ln}\ |\beta_1||$. Conversely, if $\mathrm{ln}\ |\beta_{(1,2)}|$ are both negative, the defect states are unambiguously localized towards the left. The smaller $|\beta_1|$ has a more negative logarithm, so $|\mathrm{ln}\ |\beta_1||> |\mathrm{ln}\ |\beta_2||$. The more ambiguous case is when $\mathrm{ln}\ |\beta_1|$ is negative and $\mathrm{ln}\ |\beta_2|$ positive, which we analyze below. 

It can be readily found that the values of $\beta$ for which the non-Bloch Hamiltonian Eq. \eqref{eq:HpBeta} has an imaginary eigenvalue $E=i\epsilon,\ \epsilon \in \mathbb{R}$ are given by 
\begin{equation}
  \beta_{\pm} = \frac{\delta^2-t_1^2-t_2^2-\epsilon^2 \pm \sqrt{4t_1^2(\delta^2-t_2^2)+(t_1^2+t_2^2+\epsilon^2-\delta^2) }}{2t_1(\delta+t_2)}. \label{eq:betapm}
\end{equation}

 Eq. \eqref{eq:betapm} implies that 
\begin{equation}
  \beta_1\beta_2 = \frac{t_2-\delta}{t_2+\delta} \label{eq:b1b20}
\end{equation}
regardless of the energy. In particular, this is also true at the generalized brillouin zone (GBZ) energies, at which $|\beta_1|=|\beta_2|=\sqrt{\frac{t_2-\delta}{t_2+\delta}}$, which we denote as $|\beta_{\mathrm{GBZ}}|$ for convenience. The sign of $\mathrm{ln}\ |\beta_{\mathrm{GBZ}}|$ indicates which boundary the bulk  eigenstates of the pristine SSH chain without defects are localized at by the NHSE with a negative (positive) sign corresponding to localization at the left (right) boundary. 

Meanwhile, if $(\delta^2-t_1^2-t_2^2)^2 > 4(t_1^2)(t_2^2-\delta^2)$, as is the case for the parameter sets considered here, both of the $\beta_\pm$ are real. Imposing the  further condition that $\delta^2 < t_2^2$ makes both of the $\beta_\pm$ negative with $\beta_1$ ($\beta_2$) corresponding to $\beta_+$ ($\beta_-$). Because both $\beta_\pm$ are real and have the same sign, $|\beta_1||\beta_2| = \beta_1\beta_2$, and Eq. \eqref{eq:b1b20} can be rewritten as  
\begin{equation}
  \mathrm{ln}\ |\beta_2| = 2 \mathrm{ln}\ |\beta_{\mathrm{GBZ}}|-\mathrm{ln}\ |\beta_1|. \label{eq:b1b21}
\end{equation}
By assumption, $\mathrm{ln}\ |\beta_2|$ is positive so $\mathrm{ln}\ |\beta_2|=|\mathrm{ln}\ |\beta_2||$ whereas $\mathrm{ln}\ |\beta_1|$ is negative, so $\mathrm{ln}\ |\beta_1| = -|\mathrm{ln}\ |\beta_1||$. Rearranging Eq. \eqref{eq:b1b21} gives 
\begin{equation}
  |\mathrm{ln}\ |\beta_2|| - |\mathrm{ln}\ |\beta_1|| = 2 \mathrm{ln}\ |\beta_{\mathrm{GBZ}}|.
\end{equation}
Therefore, if $\mathrm{ln}\ |\beta_{\mathrm{GBZ}}|$ is positive (negative), $|\mathrm{ln}\ |\beta_2||$ is larger (smaller) than $|\mathrm{ln}\ |\beta_1||$ and the wavefunction in a chain with multiple regularly spaced defects is localized towards the right (left) boundary in a consistent manner to the OBC NHSE localization direction of the pristine SSH chain without defects.

%\textcolor{red}{\subsection{Reversed Non-Reciprocity}}

%\textcolor{red}{Reversing \(\delta < 0\) flips the NHSE direction to the right edge. See Fig. S5 for rightward accumulation.}

%\bibliographystyle{apsrev4-2}
%\bibliography{Arxiv_version/defectskin1arxiv}


\end{document}